# Optical Frequency Comb Fourier Transform Spectroscopy with Sub-Nominal Resolution - Principles and Implementation

Lucile Rutkowski,[1] Piotr Masłowski,[2] Alexandra C. Johansson,[1] Amir Khodabakhsh,[1] and Aleksandra Foltynowicz[1,*]

[1]*Department of Physics, Umeå University, 901 87 Umeå, Sweden*
[2]*Institute of Physics, Faculty of Physics, Astronomy and Informatics, Nicolaus Copernicus University in Toruń, ul. Grudziądzka 5, 87-100 Toruń, Poland*
*Corresponding author: aleksandra.foltynowicz@umu.se*

## Abstract

Broadband precision spectroscopy is indispensable for providing high fidelity molecular parameters for spectroscopic databases. We have recently shown that mechanical Fourier transform spectrometers based on optical frequency combs can measure broadband high-resolution molecular spectra undistorted by the instrumental line shape (ILS) and with a highly precise frequency scale provided by the comb. The accurate measurement of the power of the comb modes interacting with the molecular sample was achieved by acquiring single-burst interferograms with nominal resolution precisely matched to the comb mode spacing. Here we give a full theoretical description of this sub-nominal resolution method and describe in detail the experimental and numerical steps needed to retrieve ILS-free molecular spectra, i.e. with ILS-induced distortion below the noise level. We investigate the accuracy of the transition line centers retrieved by fitting to the absorption lines measured using this method. We verify the performance by measuring an ILS-free cavity-enhanced low-pressure spectrum of the $3\nu_1+\nu_3$ band of $CO_2$ around 1575 nm with line widths narrower than the nominal resolution. We observe and quantify collisional narrowing of absorption line shape, for the first time with a comb-based spectroscopic technique. Thus retrieval of line shape parameters with accuracy not limited by the Voigt profile is now possible for entire absorption bands acquired simultaneously.

## 1. Introduction

Fourier transform spectrometers (FTS) are well-established detection systems for molecular spectroscopy because of their ability to measure broadband spectra with an absolutely calibrated frequency scale. They are commonly used in combination with incoherent light sources in Fourier transform infrared (FTIR) spectroscopy [1]. An FTS is a Michelson interferometer that produces an interferogram of the light source when the optical path difference (OPD - Δ) between the interferometer arms is scanned. Fourier transform of the interferogram yields a spectrum spanning the entire bandwidth of the analyzed light source with a nominal resolution inversely proportional to the range of the OPD scan, $\Delta_{max}$. Because of the OPD truncation, the absorption spectrum is convolved with an instrumental line shape (ILS) [2]. Therefore, high nominal resolution is needed to avoid distortion of the absorption lines in conventional FTIR spectroscopy, which can only be achieved with a large instrument [3]; for instance, a 100 MHz resolution necessitates scanning the OPD over 3 m. Furthermore, since the OPD is usually calibrated using a stabilized continuous wave (cw) laser, the accuracy of the FTS frequency scale is affected by e.g. the beam divergence, interferometer misalignment, and fluctuations of thermodynamic conditions. Achieving high accuracy on the frequency and intensity scales requires placing the entire spectrometer in vacuum and careful alignment, which are challenging when the instrument size grows.

The advent of optical frequency combs (OFC) provided a new light source for the FTS [4]. The high spectral brightness together with the spatial and temporal coherence of the combs enable acquisition times orders of magnitude shorter than in conventional FTIR spectroscopy. In the temporal domain, a comb is a train of pulses separated by $1/f_{rep}$, where $f_{rep}$ is the repetition rate, therefore a comb-based FTS interferogram consists of a series of equidistant bursts separated by $c/f_{rep}$ in the OPD domain, where c is the speed of light. The spectral comb structure starts to be visible when an interferogram containing several bursts is acquired [5]. However, unless the interferogram is sampled correctly, the spectral resolution is limited to the nominal resolution of the FTS and retrieving the power of comb modes requires interpolation between the FTS sampling points [6]. Recently, we have shown that the power of the comb modes can be precisely measured using single-burst interferograms with the length matched to $c/f_{rep}$ [7]. Such measurement yields high resolution molecular spectra free from ILS distortion even when the molecular line widths are much narrower than the nominal resolution of the spectrometer. Moreover, the frequency scale is given by the comb, so high frequency precision and accuracy are achieved even when the interferometer is not placed in vacuum.

Here we use this sub-nominal resolution method to measure an ILS-free cavity-enhanced low-pressure spectrum of the entire $3\nu_1+\nu_3$ band of $CO_2$ in ~10 minutes with frequency scale precision and signal-to-noise ratio high enough to observe the influence of the speed-dependent effects on the absorption line shape. We describe in detail the principles of the method and the different steps that are performed on a single-burst interferogram to obtain an ILS-free molecular spectrum. We discuss the influence of the uncertainty of the cw reference laser wavelength used for OPD calibration on the accuracy of the retrieved line center frequencies, and we show that the technique is particularly suited for high precision spectroscopy of narrow spectral features.

## 2. Method

### 2.1. Truncation of optical path difference: nominal resolution and instrumental line shape

The basic principle of an FTS is to acquire an interferogram of the light source when the OPD ($\Delta$) between the arms of the interferometer is scanned. In conventional FTIR spectroscopy based on incoherent light sources, the interferogram consists of a single burst occurring at $\Delta=0$. When permitted by the interferometer length, it is convenient to acquire double-sided interferograms to simplify the phase correction in post-processing [1]. The interferogram is usually sampled at OPD steps equal to a simple fraction of the cw reference laser wavelength, $\lambda_{ref}$, and the final spectrum is obtained by taking the magnitude of the fast Fourier transform (FFT) of the interferogram. In general, the spectrum is affected by an instrumental line shape (ILS) originating from the OPD truncation and the divergence and misalignment of the beams inside the spectrometer [1]. However, when the light source is a collimated laser beam, as is the case for the comb, the divergence and misalignment effects are negligible and the ILS is caused mainly by the OPD truncation. For a boxcar acquisition function in the OPD domain, the truncation-induced ILS is a normalized sinc function defined as:

$$g_{ILS}(\nu,\nu_0) = \frac{\Delta_{max}}{c} \frac{\sin\left[\pi\left(\nu-\nu_0\right)\Delta_{max}/c\right]}{\pi\left(\nu-\nu_0\right)\Delta_{max}/c}, \qquad (1)$$

where $c/\Delta_{max}$ is the nominal resolution of conventional FTIR, $\nu$ is the frequency and $\nu_0$ is the center frequency (all in Hz). The truncation-induced ILS, which has periodic zero-crossings at detunings ($\nu-\nu_0$) equal to integer multiples of $c/\Delta_{max}$, is convolved with the spectrum measured by the FTS, leading to a broadening of the measured lines, reduction of their intensity, and ringing on each side of the lines. Using apodization methods that involve a modification of the acquisition function of the interferogram [1], it is possible to smoothen the ringing resulting from the convolution with the sinc function at the expense of a further broadening of the lines.

### 2.2. Comb-based FTS

When a comb beam is injected in the FTS, the resulting interferogram is a train of bursts equidistant in the OPD domain and separated by $c/f_{rep}$. As mentioned above, the comb structure becomes visible when at least two consecutive bursts are acquired in the same interferogram, since the nominal resolution is then smaller than $f_{rep}$. In this case, the number of spectral elements in the spectrum after the FFT is at least twice higher than the number of comb modes. Moreover, without careful sampling of the interferogram a fit to the spectral pattern is needed to precisely find the power of each comb mode [5, 8]. Therefore, the ideal case is to measure a spectrum containing one sampling point per comb mode. As previously demonstrated in [7], it is sufficient to acquire a single-burst interferogram with a length matched to $c/f_{rep}$ to measure the power of all comb modes with high precision. Such single-burst interferogram, centered on a burst and

sampled with an OPD step equal to $\lambda_{ref}/q$, where q is an integer, contains a number of points:

$$N_0 = \text{round}\left(q\frac{c}{2\lambda_{ref} f_{rep}}\right) \qquad (2)$$

on each side of the burst. The spectrum after FFT contains $N_0$ points at frequencies ranging from 0 to $qc/2\lambda_{ref}$ (the Nyquist frequency) spaced by the nominal resolution of a conventional FTIR spectrometer:

$$f_{FTS}^0 = \frac{c}{\Delta_{max}} = \frac{q\,c}{2\lambda_{ref}\,N_0}. \qquad (3)$$

The FTS sampling points can be indexed by an integer number n and yield a frequency scale:

$$\nu_{FTS} = n\,f_{FTS}^0. \qquad (4)$$

Meanwhile, the frequency scale of the comb is given by:

$$\nu_{OFC} = n\,f_{rep} + f_{ceo}, \qquad (5)$$

where $f_{ceo}$ is the carrier-envelope offset frequency. The line width of the modes of a stabilized comb is much smaller than the width of the truncation-induced ILS, therefore the comb modes can be considered as Dirac $\delta$-functions. Thus the convolution of the comb spectrum with the truncation-induced ILS simplifies to a product, and the FTS spectrum after FFT of a single-burst interferogram is proportional to the sum of the comb mode powers, P, multiplied by their corresponding truncation-induced ILS:

$$S_{FTS}(\nu) = A\sum_n P_n g_{ILS}\left(\nu, \nu_{OFC}^n\right), \qquad (6)$$

where A is an instrumentational factor that contains the detector responsivity, gains, etc. At the FTS sampling points this becomes:

$$S_{FTS}\left(\nu_{FTS}^n\right) = A\sum_m \frac{P_m}{f_{FTS}^0} \frac{\sin\left[\pi\left(\nu_{FTS}^n - \nu_{OFC}^m\right)/f_{FTS}^0\right]}{\pi\left(\nu_{FTS}^n - \nu_{OFC}^m\right)/f_{FTS}^0}, \qquad (7)$$

where Eqs (1) and (3) have been used to express the ILS in terms of the FTS sampling point spacing.

The influence of the ILS and the FTS sampling on a comb spectrum is illustrated in Fig. 1, which shows the spectrum of a single-burst interferogram whose length is longer than $c/f_{rep}$ [panel (a), in red] and exactly equal to $c/f_{rep}$ [panel (b), in blue]. For simplicity, $f_{ceo}$ is assumed to be null. The comb modes are represented by the black $\delta$-functions, and the central comb mode, $n_{abs}$, is absorbed by a molecular transition with a line width much narrower than $f_{rep}$. The dashed curves show the truncation-induced ILS of the central comb mode, and similar ILS (not shown) exists for all comb modes. The FTS spectrum resulting from an addition of the ILS of all comb modes, shown by the solid curves, is sampled by the FTS at discrete points separated by $f_{FTS}^0$ (circular markers, dotted curves are guides to the eye). When the nominal resolution is smaller than $f_{rep}$ [panel (a)], the FTS spectrum at the mode positions is not proportional to the comb mode power because of the distortion caused by the ILS of the neighboring comb modes. Moreover, the spacing of the sampling points is smaller than the spacing of the comb modes. This mismatch accumulates with n and results in a substantial frequency shift between the FTS sampling points and the comb frequencies, which causes the ringing on each side of the absorbed mode in the final spectrum. The influence of the ILS vanishes when the nominal resolution is equal to $f_{rep}$ [panel (b)]. In this case, the zero-crossings of the ILS of one comb mode are at the positions of the neighboring comb modes, leaving them unaffected. The sum of the ILS is proportional to the comb mode power, and the sampling points, separated by $f_{rep}$, are precisely matched to the comb mode frequencies, so the final spectrum [markers in Fig. 1(b)] has a flat baseline on each side of the absorbed mode.

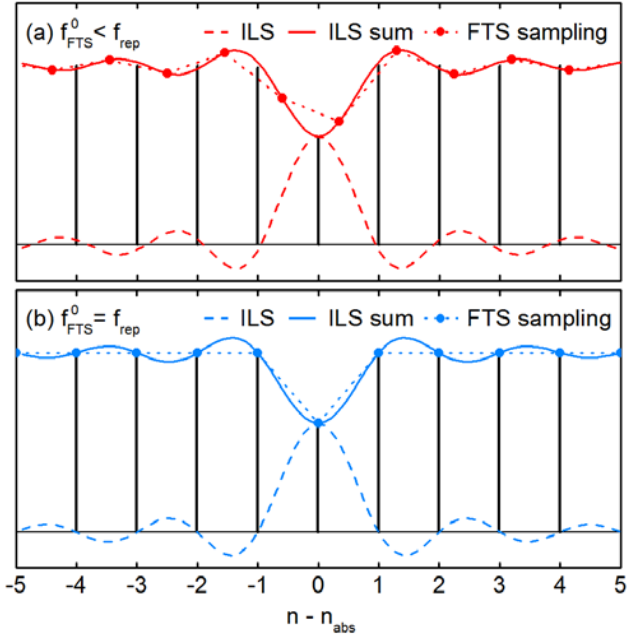

Fig. 1. Schematics of the spectra obtained after FFT of a single-burst interferogram when the nominal resolution is smaller than $f_{rep}$ (a) and exactly equal to $f_{rep}$ (b). The comb modes are symbolized by the black vertical lines. The dashed red (a) and blue (b) curves are the ILS profiles of the absorbed central comb mode, $n_{abs}$. The solid red (a) and blue (b) curves show the FTS spectrum resulting from the summation of the ILS profiles of all comb modes. The frequencies sampled by the FTS are marked with red (a) and blue (b) circular markers, the dotted lines are guides to the eye.

In practice, it is not always possible to precisely match the nominal resolution of the spectrometer to $c/f_{rep}$, since $f_{FTS}^0$ can only take discrete values determined by the product of $\lambda_{ref}/q$ and the integer number of points $N_0$, while $f_{rep}$ can take any value from the laser tuning range. The smallest possible step for $f_{FTS}^0$, obtained from Eq. (3) by changing the interferogram length by two points, is $qc/[2N_0 (N_0+1) \lambda_{ref}]$. This yields 594 Hz for $\lambda_{ref} = 633$ nm, $q = 4$, and $f_{rep} = 750$ MHz, i.e. $N_0 = 1.26 \times 10^6$. Stepping $f_{rep}$ by this value results in an optical shift of the comb modes by 150 MHz at 1575 nm, which is too large to map e.g. the absorption profiles of narrow molecular transitions. Therefore in practice $f_{rep}$ has to be tuned by a smaller step than the available $f_{FTS}^0$ step and the two do not always match. This implies that in general there is a relative discrepancy between $f_{FTS}^0$ and $f_{rep}$:

$$\varepsilon_0 = \frac{f_{FTS}^0 - f_{rep}}{f_{rep}}, \qquad (8)$$

which takes a value between 0 and $1/2N_0$, the latter for the case when the $f_{rep}$ value lies exactly in between two available $f_{FTS}^0$ values. The maximum discrepancy is thus of the order of $10^{-6}$ for $N_0$ of the order of $10^6$. For such a small difference between $f_{FTS}^0$ and $f_{rep}$ the relative distortion of the FTS spectrum at the comb mode positions is of the same order as $\varepsilon_0$, which is negligible (see section 2.5 for detailed discussion). However, this discrepancy also causes a significant shift of the sampling points with respect to the comb line positions, as the small mismatch between $f_{FTS}^0$ and $f_{rep}$ is multiplied by the comb mode number n, which is of the order of $10^5$. Moreover, the FTS scale does not exhibit any offset frequency while $f_{ceo}$ can take any value between 0 and $f_{rep}$. Therefore in the following sections we assume that $f_{FTS}^0$ is chosen closest possible to $f_{rep}$ and we describe how to correct the FTS frequency scale in order to sample the spectrum at the positions of the comb lines.

### 2.3. Matching the FTS and OFC frequency scales

As described above, to sample the power of the comb modes correctly, the FTS scale [Eq. (4)] has to be equal to the OFC scale [Eq. (5)]. However, in the experiment these two scales most often do not match, therefore a numerical procedure, illustrated in Fig. 2, is needed to correct the FTS scale in post-processing. The OFC scale (red solid line) is set by the experiment. The initial FTS scale (blue solid line), set by the cw reference laser wavelength and the number of points in the interferogram, has in general a different slope and no offset. The offset is corrected by shifting the FTS scale by $f_{ceo}$ (blue dotted line) and the remaining discrepancy between the frequency scales is then equal to the error on the FTS sampling point spacing multiplied by the index n of the considered mode. This error can be corrected at a chosen mode position $n_{opt}$ (corresponding to the optical frequency $\nu_{opt}$ within the comb spectrum) by an additional frequency shift of $f_{shift}^0$ (blue dash-dotted line). In this way, the error on the sampling point spacing remains but is now multiplied by a much smaller number ($n-n_{opt}$). When matching of the scales is required over a wider spectral range, the FTS sampling point spacing, and thus the slope of the FTS scale, can be corrected through spectral interpolation by zero-padding of the interferogram, and afterwards, if needed, a smaller frequency shift, $f_{shift}$, can be performed (not indicated in the figure) to cross the two scales at the desired frequency (blue dashed line).

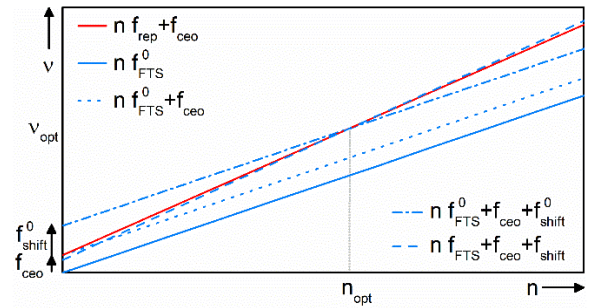

Fig. 2. Schematics of the process of matching the FTS scale (blue solid line) to the OFC scale (red solid line) in the spectral range of interest (around $\nu_{opt}$). For clarity, the initial mismatch is exaggerated and all scales are shown as continuous functions rather than discrete points. In the first step, $f_{ceo}$ is added to the FTS scale so that both scales have the same frequency origin (blue dotted line). In the second step, the FTS scale is shifted by $f_{shift}^0$ so that it overlaps with the OFC scale in a chosen optical range of interest (blue dash-dotted line), or the interferogram is zero-padded to correct the slope of the FTS scale before the shift is performed (blue dashed line).

### 2.3.1. Offset correction

The FTS offset frequency is null by definition of the Fourier transform process while the comb offset frequency can take any value between 0 and $f_{rep}$. Although techniques exist to stabilize the comb $f_{ceo}$ to zero [9], it is most often locked to a frequency different than zero for simplicity or for practical reasons (e.g. when the comb is coupled to an optical cavity that does not permit arbitrary tuning of the $f_{ceo}$). Therefore the FTS scale has to be shifted by $f_{ceo}$, which is done by performing FFT of the interferogram power $P(\Delta)$ multiplied by an exponential function containing $f_{ceo}$:

$$S_{FTS}\left(\nu_{FTS}\right) = A\,abs\left\{FFT\left[P(\Delta)\exp\left(-i2\pi f_{ceo}\,\Delta/c\right)\right]\right\}. \quad (9)$$

As a result, the FTS scale becomes $\nu_{FTS} = n\,f^0_{FTS} + f_{ceo}$, which corresponds to the blue dotted FTS scale in Fig. 2, and the spectrum is recalculated at the new frequencies.

### 2.3.2. FTS sampling point spacing correction

Because of the discrete sampling of the interferogram $f^0_{FTS}$ can only take discrete values determined by the product of $\lambda_{ref}/q$ and $N_0$, while $f_{rep}$ can be continuously tuned within the range allowed by the laser actuators, and the two cannot always be precisely matched. This introduces an error between the $n^{th}$ comb line frequency and its corresponding sampling point given by $n\varepsilon_0 f_{rep}$. After performing the $f_{ceo}$ shift, a local agreement of both scales can be achieved at index $n_{opt}$ by shifting the FTS scale by $f^0_{shift}$ given by:

$$f^0_{shift} = -\,n_{opt}\left(f^0_{FTS} - f_{rep}\right) = -\,n_{opt}\,\varepsilon_0\,f_{rep}. \quad (10)$$

This shift is performed in the same way as the $f_{ceo}$ shift [Eq. (9)], and yields the dash-dotted FTS scale in Fig. 2. This is a simple and computationally efficient way to equalize the two scales at index $n_{opt}$. Moreover, it reduces the error between the other comb lines and their corresponding FTS sampling points to $(n-n_{opt})\varepsilon_0 f_{rep}$. However, if the spectrum spans a large bandwidth (thousands of comb lines), this error might become significant away from $n_{opt}$. In such a case, the discrepancy between the FTS sampling point spacing and $f_{rep}$ can be decreased by zero-padding of the interferogram. This spectral interpolation method involves the addition of $k_{pad}N_0$ zeros (with integer $k_{pad}$) on both sides of the existing interferogram before performing the FFT. The resulting spectrum contains $k_{pad}$ interpolated points between two previously sampled points, spaced by $f^0_{FTS}/(k_{pad}+1)$. If instead the number of zeros added on both sides is close to but not equal to $k_{pad}N_0$, the FTS sampling point spacing changes. The number of points in the zero-padded interferogram, N, that shifts the sampling points closest to the comb modes for a given integer $k_{pad}$ is obtained by replacing $f_{rep}$ in Eq. (2) by $f_{rep}/(k_{pad}+1)$:

$$N = round\left[\frac{q\,c}{2\lambda_{ref}\,f_{rep}}\left(k_{pad}+1\right)\right]. \quad (11)$$

The new FTS sampling point spacing is:

$$f_{FTS} = \frac{q\,c}{2\lambda_{ref}\,N}\left(k_{pad}+1\right), \quad (12)$$

yielding a relative change with respect to the initial FTS sampling point spacing of:

$$\zeta = \frac{f_{FTS} - f^0_{FTS}}{f_{FTS}}. \quad (13)$$

Note that the maximum value $\zeta$ can take is $\varepsilon_0$ (when $f_{FTS} = f_{rep}$), which, in turn, is at most $1/2N_0$. The relative discrepancy $\varepsilon$ between $f_{FTS}$ and $f_{rep}$ is now:

$$\varepsilon = \frac{f_{FTS} - f_{rep}}{f_{rep}}. \quad (14)$$

The smaller $\varepsilon$ is, the more parallel the red OFC and blue dashed FTS scales are in Fig. 2. The optimum number $k_{pad}$ that decreases the discrepancy between the sampling point spacing and $f_{rep}$ without over-padding of the interferogram can be found by evaluating $\varepsilon$ as a function of $k_{pad}$. Figure 3 shows examples of curves obtained when calculating $\varepsilon(k_{pad})$ for q = 4, $f^0_{FTS}$ = 750 MHz and three different values of $f_{rep}$, off by 40, 60 and 100 Hz from 750 MHz. The smaller the initial discrepancy $\varepsilon_0$ is, the larger $k_{pad}$ is needed to significantly reduce $\varepsilon$.

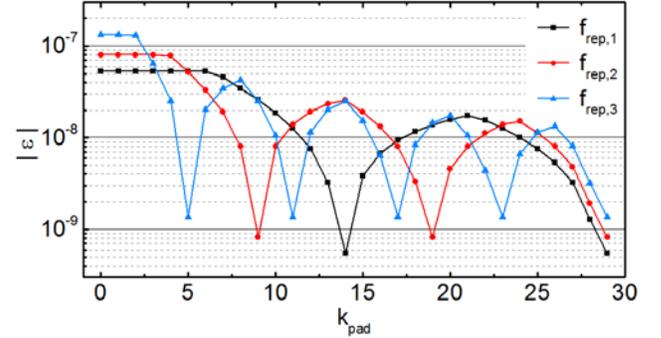

Fig. 3. Relative discrepancy of the FTS sampling point spacing with respect to the target $f_{rep}$ value as a function of $k_{pad}$ for different $f_{rep}$ values. $f^0_{FTS}$ is set to 750 MHz, q to 4, while $f_{rep}$ is detuned from 750 MHz by 40 Hz ($f_{rep,1}$, black square markers), 60 Hz ($f_{rep,2}$, red circular markers) and 100 Hz ($f_{rep,3}$, blue triangular markers, the curves are guides to the eye).

As computational time and memory requirement increase with the number of points involved in the FFT algorithm, it is computationally more efficient to perform zero-padding with a low $k_{pad}$ and combine it with the shift procedure, which does not increase the number of points. The shift is then given by:

$$f_{shift} = -\,n_{opt}\left(f_{FTS} - f_{rep}\right) = -n_{opt}\,\varepsilon f_{rep}, \quad (15)$$

and the FTS frequency scale becomes (blue dashed FTS scale in Fig. 2):

$$\begin{aligned}\nu^n_{FTS} &= n\,f_{FTS} + f_{ceo} + f_{shift}\\ &= n\,f_{rep} + f_{ceo} + \left(n - n_{opt}\right)\varepsilon f_{rep},\end{aligned} \quad (16)$$

where Eqs (14) and (15) have been used in the last step. This means that the discrepancy between the FTS and OFC scales is given by $(n-n_{opt})\varepsilon f_{rep}$. Once the offset and sampling point spacing corrections are done the comb mode frequencies are assigned to the points sampled by the FTS.

### 2.4. Interleaving the spectra

When a molecular absorption line is much narrower than $f_{rep}$, only a single comb mode is strongly absorbed by the transition, resulting in one sampling point per absorption line. Mapping the full profile of absorption lines thus requires interleaving of single-burst spectra taken with different OFC scales [10], which is achieved by stepping either $f_{rep}$ or $f_{ceo}$. Stepping $f_{ceo}$ is preferred, as it results in a linear shift of the optical frequencies, which is easily taken care of in post-processing. However, this is not possible in e.g. cavity-enhanced techniques, in which the comb parameters must be locked to that of an external cavity. In this case $f_{rep}$ is stepped instead, together with the cavity free spectral range.

### 2.5. Residual ILS distortion

The procedure described in section 2.3 assumes that $f^0_{FTS}$ is known with the same accuracy and precision as $f_{rep}$. In practice, $f_{rep}$ is usually stabilized and known with accuracy on the $10^{-11}$ level, while the accuracy of $f^0_{FTS}$ is much lower, determined by the accuracy $\eta$ with which $\lambda_{ref}$ is known, limited by the variations of the refractive index of air (when the FTS is not in vacuum) and of the alignment of the laser beam [1]. Thus in practice the cw reference laser wavelength is given by $\lambda'_{ref} = \lambda_{ref}(1+\eta)$. This implies that relative discrepancy between the corrected FTS sampling point spacing and $f_{rep}$ [Eq. (14)] becomes $\varepsilon = f_{FTS}/[(1+\eta)\ f_{rep}] - 1$, which implies that $f_{FTS} = (1+\varepsilon)(1+\eta)\ f_{rep} \approx (1+\varepsilon+\eta)\ f_{rep}$, where the term of the order of $\varepsilon\eta$ has been neglected. The corrected FTS frequency scale thus becomes:

$$\nu^n_{FTS} = nf_{rep}\left(1 + \varepsilon + \eta\right) + f_{ceo} - n_{opt}\varepsilon f_{rep}$$
$$= nf_{rep} + f_{ceo} + \left[\left(n - n_{opt}\right)\varepsilon + n\eta\right]f_{rep}. \qquad (17)$$

Below we show that the residual mismatch between the FTS and OFC scales causes a characteristic residual ILS distortion with amplitude proportional to the error remaining between them, which, in turn, allows iterative reduction of $\eta$ until the ILS distortion becomes lower than the noise on the baseline. We also study the influence of error $\eta$ on the position of absorption lines retrieved by fitting and show that it depends on the ratio of the absorption line width, $\Gamma$, and $f_{rep}$.

Using Eq. (7) the FTS spectrum at the $n^{th}$ sampling point after correction of the FTS scale can be written as:

$$S_{FTS}\left(\nu^n_{FTS}\right) =$$
$$\frac{A}{f^0_{FTS}}\sum_m P_m \frac{\sin\left[\pi(n-m)f_{FTS}/f^0_{FTS} + \pi\left(\nu^m_{FTS}-\nu^m_{OFC}\right)/f^0_{FTS}\right]}{\pi(n-m)f_{FTS}/f^0_{FTS} + \pi\left(\nu^m_{FTS}-\nu^m_{OFC}\right)/f^0_{FTS}}, \qquad (18)$$

where $\nu^n_{FTS}-\nu^m_{FTS} = (n-m)\ f_{FTS}$ has been used. Since $f_{FTS}/f^0_{FTS} \approx 1$ and $\nu^m_{FTS}-\nu^m_{OFC} \ll f^0_{FTS}$, the sine function can be series expanded to first order and Eq.(18) becomes:

$$S_{FTS}\left(\nu^n_{FTS}\right) \approx$$
$$\frac{A}{f^0_{FTS}}\sum_m P_m \left(-1\right)^{n-m}\frac{(n-m)\left(f_{FTS}-f^0_{FTS}\right) + \nu^m_{FTS}-\nu^m_{OFC}}{(n-m)f_{FTS} + \nu^m_{FTS}-\nu^m_{OFC}}, \qquad (19)$$

where $\left(\nu^m_{FTS}-\nu^m_{OFC}\right) \ll (n-m)f_{FTS}$ for $n \neq m$, which enables using Eq. (13) to further simplify the equation to:

$$S_{FTS}\left(\nu^n_{FTS}\right) \approx \frac{A}{f^0_{FTS}}\left\{P_n + \sum_{m \neq n} P_m \left(-1\right)^{n-m}\left[\zeta + \frac{\nu^m_{FTS}-\nu^m_{OFC}}{(n-m)f_{FTS}}\right]\right\}. \qquad (20)$$

In the absence of molecular absorption the comb mode power is locally flat. This implies that $S_{FTS}(\nu^n_{FTS}) \approx A\,P_n/f^0_{FTS}$ because the contributions of neighboring comb modes cancel out after summation. When a comb mode $n_{abs}$ is absorbed by a molecular line with a width narrower than $f_{rep}$, the comb power changes by $\Delta P_{abs}$ at this mode, while it remains constant at the neighboring modes, $\Delta P_{n \neq n_{abs}} = 0$. This induces an ILS distortion, $\Delta S_{ILS}$, around this mode, i.e. at sampling points $n \neq n_{abs}$, given by the last term in Eq. (20) evaluated for $n = n_{abs}$:

$$\Delta S_{ILS}\left(\nu^{n \neq n_{abs}}_{FTS}\right)$$
$$\approx \frac{A}{f^0_{FTS}}\Delta P_{n_{abs}}\left(-1\right)^{n-n_{abs}}\left[\zeta + \frac{\left(n_{abs} - n_{opt}\right)\varepsilon + n_{abs}\eta}{n - n_{abs}}\right], \qquad (21)$$

where Eq. (17) has been used to express the difference between the FTS and OFC scales, and $f_{rep}/f_{FTS} \approx 1$ has been assumed. The spectrum is ILS-free when the maximal ILS distortion is smaller than the noise on the baseline, i.e. when the last term in Eq. (21) is smaller than the inverse of the signal-to-noise ratio (SNR) of the measured absorption line. The contribution of $\zeta$ to the residual ILS is negligible as long as $\zeta <$

1/SNR, which is fulfilled as long as the relative noise is larger than $10^{-6}$. The influence of the inaccuracy of the FTS sampling point spacing, $\varepsilon$, can always be removed for a particular absorption line by setting $n_{opt} = n_{abs}$ in the shifting procedure. Therefore, in practice, the residual ILS distortion is determined by the error on the reference laser wavelength, $\eta$. This distortion has odd symmetry with respect to the absorption line center since it changes sign periodically with $(n-n_{abs})$, and a maximum at $f_{rep}$ away from the absorption line center, i.e. at the comb mode $n = n_{abs} \pm 1$. An ILS-free spectrum is thus obtained when $\lambda_{ref}$ is known with accuracy better than the limit set by the SNR in the spectrum, given by:

$$\eta_{lim} = \frac{1}{SNR\ n_{abs}}. \qquad (22)$$

This limit is of the order of $1\times10^{-8}$ for a comb mode at 1575 nm and $f_{rep}$ = 750 MHz (i.e. $n_{abs} = 2.5\times10^5$) and an absorption line measured with a SNR of 400.

Equation (21) gives an analytical expression for the residual ILS distortion of an absorption line significantly narrower than $f_{rep}$. To illustrate how the residual ILS distortion depends on the ratio of absorption line width, $\Gamma$, and $f_{rep}$, three different cases are plotted in Fig. 4, for $\Gamma/f_{rep}$ equal to 0.1 [panel (a)], 1 [panel (b)] and 3 [panel (c)]. The absorption lines are calculated using Beer-Lambert law and Voigt profiles with a Doppler width equal to twice the Lorentzian width and with an absorption line contrast of 10%. The index of the absorbed comb mode closest to resonance is $n_{abs} = 2.54\times10^5$. Single-burst interferograms are simulated using a fixed value of $\lambda_{ref}$ for different $f_{rep}$ values and analyzed with $\lambda'_{ref} = \lambda_{ref}(1+\eta)$, with $\eta = 1\times10^{-7}$. $f_{rep}$ is stepped by $\Gamma/20/n_{abs}$, yielding 20 points per $\Gamma$ after interleaving. The resulting interleaved spectra are shown with black markers in Fig. 4. The red curves show fits of the model based on the Voigt profile with line position, area and Lorentzian width as fitting parameters, and the residuals of the fits are plotted in the lower panels. When the absorption line is significantly narrower than $f_{rep}$ [panel (a)] only one comb mode is absorbed by the line in each single-burst spectrum and the ILS distortion has a form of discrete peaks at multiples of $f_{rep}$ away from the absorption line center, with amplitude changing sign periodically. Since these peaks are clearly separated from the absorption line and their combined area is zero, they do not distort the absorption line. The fit is not affected by the ILS distortion, it returns the line parameter values assumed in the simulation and the residual is flat at the position of the absorption line. The residual also confirms that the maximum amplitude of the ILS distortion is equal to $n_{abs}\eta\,\Delta P_{nabs} = 2.54\times10^{-3}$, as predicted by Eq. (21). When the line width is comparable to $f_{rep}$ [panel (b)] several comb modes are simultaneously absorbed in each single-burst spectrum and the ILS distortion takes the form of ringing with odd symmetry. This is because the different discrete ILS peaks centered at multiples of $f_{rep}$ away from the line center are now broader and merge. The area and line width are not affected by the distortion, since it has odd symmetry and its integral is zero, but the center frequency of the line is shifted by -0.65$\eta n_{abs}f_{rep}$. Finally, in the case when $\Gamma/f_{rep}$ = 3 [panel (c)], the ILS ringing becomes negligibly small and the fit returns the correct area and line width. However, the center frequency of the line is shifted by -$\eta n_{abs}f_{rep}$, similar to what happens in conventional FTIR spectroscopy with an error $\eta$ on the reference laser wavelength [1].

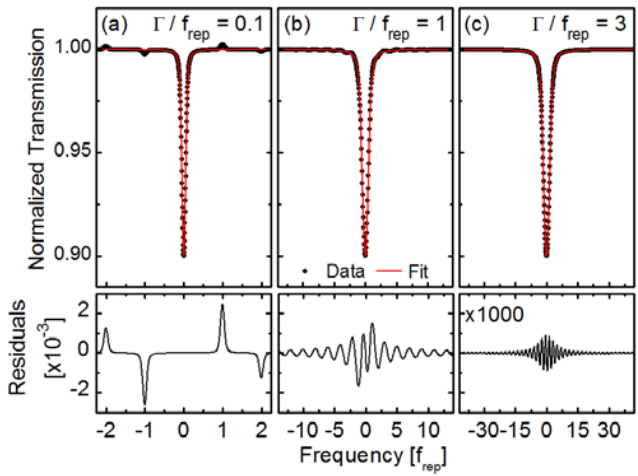

Fig. 4. Simulated absorption lines (black markers) with residual ILS distortion caused by an error $\eta = 1 \times 10^{-7}$ on the cw reference laser wavelength for absorption line widths equal to $f_{rep}/10$ (a), $f_{rep}$ (b) and $3f_{rep}$ (c). The red curves show the fits, and residuals are plotted in the lower panels.

The characteristic shape of the ILS distortion in the sub-nominal method, which is different than the truncation-induced ILS of conventional FTIR or higher order corrections to the absorption line shape, allows performing corrections to the cw reference laser wavelength. The amplitude and sign of the ILS distortion are proportional to the magnitude and sign of $\eta$, so the initially assumed $\lambda'_{ref}$ can be corrected by repeating the data analysis in an iterative manner until the ILS disappears below the noise on the baseline. In this way $\eta$ is reduced down to the value $\eta_{lim}$ determined by the SNR in the spectrum and the absorbed comb mode number [Eq. (22)]. For lines broader than $f_{rep}$ the shift of the line position remains, however reduced by a factor $\eta/\eta_{lim}$ compared to its initial value.

Note that if initially $\eta$ is larger than $1/2n_{abs}$, the iterative matching of the FTS and OFC scales performed as described above will result in a spectrum with no ILS distortion but frequency shifted by an integer multiple of $f_{rep}$, see Eq. (17). This shift, however, is easily noticed by comparison with positions of known absorption lines, and can be eliminated.

### 2.6. Absorption line positions

The origin of the frequency shift of the absorption line position described above has its explanation in the sampling process of the sub-nominal resolution method. Figure 5 depicts the influence of the error $\eta$ on the spectrum of an absorption line for $\Gamma/f_{rep}$ equal to 0.1 [panel (a)] and 2 [panel (b)]. The index of the absorbed comb mode $n_{abs}$ is $2.54 \times 10^5$, $\eta$ is equal to $-5 \times 10^{-7}$, and $n_{opt} = n_{abs}$ to cancel the influence of $\varepsilon$ in Eq. (21). The comb modes are symbolized by the black vertical lines. Some comb modes are partially absorbed by an absorption line plotted with the black dashed curves. The red solid curves depict the sum of the ILS profiles of all comb modes [Eq. (18)] and the red circular markers are the FTS sampling points after scale correction, offset from the comb modes by $-\eta n_{abs} f_{rep}$ because of the error $\eta$ on the reference laser wavelength [Eq. (17)]. The blue square markers show the FTS sampling points after the comb mode frequencies have been assigned to them. The final spectra, obtained after interleaving of many single-burst spectra, are shown by the blue solid curves. In the case of a line narrower than $f_{rep}$ [panel (a)], the ILS sum in a single-burst spectrum is much broader than the absorption line and the amplitude error induced by $\eta$ is negligibly small, e.g. below $10^{-3}$ for $\eta$ below $10^{-7}$. This means that after assigning the comb mode frequencies to the FTS sampling points the interleaved spectrum (blue curve) overlaps with the absorption line. In the case of a line significantly wider than $f_{rep}$ [panel (b)] the ILS distortion has a negligible impact on the absorption line intensity and the ILS sum overlaps with the absorption line. Yet, once the comb mode frequencies are assigned to the FTS sampling points, the absorption line is shifted by $-\eta n_{abs} f_{rep}$.

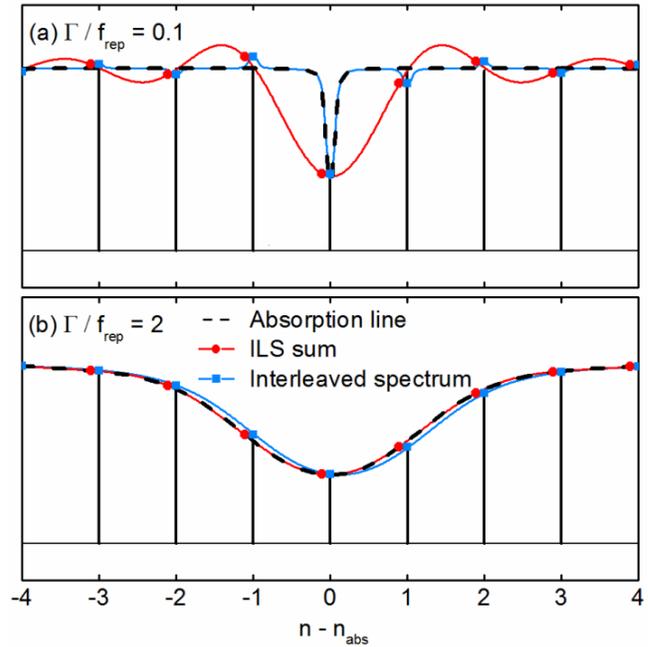

Fig. 5. Simulation of the spectra of absorption lines with line width equal to $f_{rep}/10$ (a) and $2f_{rep}$ (b) obtained when $\eta = -5 \times 10^{-7}$. The comb modes are symbolized by the black vertical lines. The absorption lines are depicted by the black dashed curves. The red solid curves are the sum of the ILS of all comb lines and the red circular markers are the FTS sampling points after scale correction. The blue square markers show the spectrum obtained after assigning the closest comb mode frequency to each FTS point, and the solid blue curve is the final spectrum, obtained after interleaving of many single-burst spectra.

For $\Gamma/f_{rep}$ ratios between the two cases shown in Fig. 5, the frequency shift of the absorption line depends on the exact value of the $\Gamma/f_{rep}$ ratio and the shape of the absorption profile. The odd symmetry of the residual ILS distortion counteracts the shift induced by the error on the reference laser wavelength because the slope of the distortion at the line center has the same sign as $\eta$. Figure 6 shows the frequency shift of the absorption line as a function of the $\Gamma/f_{rep}$ ratio, obtained by simulations and fits similar to the ones shown in Fig. 4. To illustrate the variation with transmission line shape, four different cases are plotted here, direct transmission, i.e. Beer-Lambert law, with 10% (solid red curve) and 50% (solid blue curve) contrast, and cavity-enhanced model [11] with the same contrasts (red and blue dashed curves, respectively), all based on Voigt profiles. For $\Gamma/f_{rep}$ ratios below 0.2 the frequency shift is close to zero, while for ratios larger than 2 it approaches the FTIR limit of $-\eta n_{abs} f_{rep}$ (indicated with the horizontal gray dashed line). This shift is equal to 1.9 MHz for the index of the absorbed comb mode $n_{abs} = 2.54 \times 10^5$, $f_{rep} = 750$ MHz and $\eta = 1 \times 10^{-8}$. For intermediate $\Gamma/f_{rep}$ ratios the exact value of the frequency shift depends on the profile of the line measured in transmission. The shift increases with absorption line contrast, since the line profile in transmission gets broader (i.e. the blue curves in Fig. 6 are shifted to the left with respect to the red curves). Moreover, for the same absorption line contrast the shift is larger in cavity-enhanced spectra than in direct absorption, because of the larger broadening (i.e. the dashed curves are shifted to the left with respect to the solid curves). In the experiment, the transition frequency uncertainty when the ILS distortion is at the noise level can be estimated using Fig. 6 for the pertinent $\Gamma/f_{rep}$ ratio and contrast of the measured absorption line.

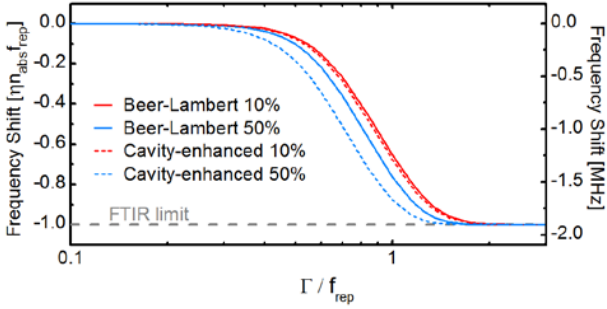

Fig. 6. Simulated shift of the transition frequency obtained by fitting to absorption lines analyzed with an error $\eta$ on the cw laser reference wavelength in direct transmission (solid curves) and in cavity transmission (dashed curves) for two absorption line contrasts, 10% (red curves) and 50% (blue curves), as a function of the ratio of the transition line width to $f_{rep}$. The y-axis on the left side is expressed in terms of the FTIR shift $\eta n_{abs} f_{rep}$, while the y-axis on the right side is in MHz for $f_{rep} = 750$ MHz, $n_{abs} = 2.54 \times 10^5$ and $\eta = 10^{-8}$.

## 3. Experimental

### 3.1. Experimental Setup

To verify the formalism and procedures described above we performed measurements using a near-infrared comb-based FTS with an enhancement cavity. The cavity was used both to increase the absorption sensitivity and to filter the comb to increase the $f_{rep}$, in order to reduce the uncertainty on the line positions (see Fig. 6). The experimental setup is depicted in Fig. 7. The comb source was an amplified femtosecond Er:fiber laser with a repetition rate of 250 MHz, equipped with an f-2f interferometer. The comb was coupled into an 80 cm long cavity with a finesse of ~11000 and a free spectral range (FSR) of 187.5 MHz. In this configuration, every third comb mode was transmitted by every fourth cavity resonance, which led to an effective (cavity-filtered) repetition rate of $f_{rep} = 750$ MHz, as shown at the bottom of Fig. 7. The wavelength-dependent cavity finesse was characterized by ring-down measurements with a precision of 1%. One of the cavity mirrors was mounted on a ring piezoelectric transducer (PZT) to control the cavity length. The cavity was connected to a supply of 1000(2) ppm of $CO_2$ diluted in $N_2$ and of pure $N_2$ through a system of flow and pressure controllers that provided 1.3 Torr precision on the sample pressure.

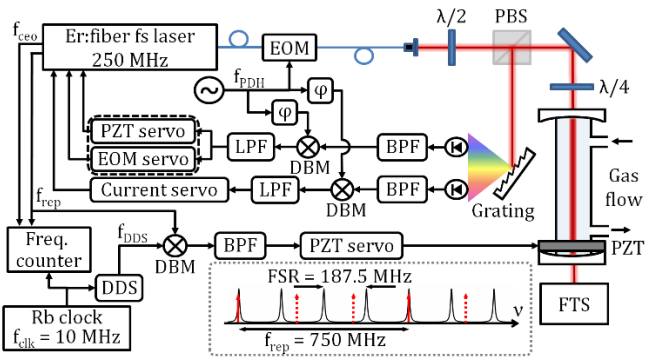

Fig. 7. Experimental setup: EOM, electro-optic modulator; $\lambda/2$, half-waveplate; PBS, polarizing beam splitter; $\lambda/4$, quarter-waveplate; PZT, piezo-electric transducer; FTS, Fourier transform spectrometer; $f_{PDH}$, Pound-Drever-Hall modulation frequency; $\varphi$, phase shifter; DBM, double-balanced mixer; BPF, band-pass filter; LPF, low-pass filter; $f_{Rb}$, Rubidium clock frequency; DDS, digital direct synthesizer; $f_{DDS}$, DDS frequency. Bottom: matching of the comb (red) to the cavity (black), the reflected comb modes are shown with dotted lines.

The comb modes were actively stabilized to the cavity resonances using the two-point Pound-Drever-Hall (PDH) technique [11]. To obtain the PDH error signals, the comb light was phase-modulated at a frequency $f_{PDH} = 20$ MHz using a fiber-coupled electro-optic modulator (EOM). The cavity reflected light was picked up and directed onto a grating that dispersed the comb spectrum. Two different wavelength regions of the dispersed light were detected by two photodetectors and their signals were synchronously demodulated at $f_{PDH}$ to yield two PDH error signals. The first error signal was fed into a proportional-integral (PI) controller that acted on the injection current of the pump diode of the oscillator and controlled both $f_{ceo}$ and $f_{rep}$ with a bandwidth of 150 kHz. The second error signal was split into two paths and fed to two separate PI controllers that worked in parallel to control $f_{rep}$. The first controller acted on the laser resonator length via an intracavity PZT, controlling $f_{rep}$ with a low bandwidth (6 kHz) but a large range (5 kHz). The second one was connected to an EOM inside the laser resonator, which modulated the intracavity refractive index and thus controlled $f_{rep}$ with a high bandwidth (500 kHz) but a small range (0.1 Hz). Once the comb modes were locked to the cavity, the cavity length/FSR was stabilized using an error signal obtained by comparing the $f_{rep}$ value to a radio-frequency sinewave, $f_{DDS}$, generated by a direct digital synthesizer (DDS) referenced to a GPS-referenced Rubidium clock. The error signal was fed via a low bandwidth (<10 Hz) PI controller to the PZT controlling the cavity length. As a consequence, $f_{rep}$ was stabilized to the Rubidium clock, while the $f_{ceo}$ was indirectly stabilized via the two-point lock to the cavity.

The light transmitted through the cavity was analyzed with a homebuilt fast-scanning FTS enclosed in a box filled with ambient air. The OPD was swept at a speed of 80 cm/s and it was calibrated with a sinewave interferogram of a stabilized He-Ne cw laser [specified $\lambda_{ref} =$ 632.9911 nm in vacuum] that co-propagated with the comb beam. The comb interferogram was measured using an autobalancing InGaAs detector [12], while the cw interferogram was measured with a single Si detector. Both interferograms were acquired simultaneously at a rate of 5 Msample/s with a 20-bit resolution data acquisition card. The comb interferogram was linearly interpolated at the zero-crossings and extrema positions of the reference interferogram during post-processing, yielding an OPD step of $\lambda_{ref}/4$ [i.e. q=4 in Eq. (2) and following]. A scan over a single-burst interferogram ($\Delta_{max} = 40$ cm) lasted 0.5 s, and the measurement could be repeated every 1.3 s.

### 3.2. Measurement procedure and comb parameters

For sub-nominal resolution measurements the cavity was filled with 1000(2) ppm of $CO_2$ in $N_2$ at a pressure of 26.3(1.3) Torr and a temperature of 296(3) K. The average full width at half maximum (FWHM) of the absorption lines at this pressure is $\Gamma = 390$ MHz. Single-burst spectra were taken with 40 different $f_{rep}$ values stepped by 75 Hz, which resulted in a step of 18.75 MHz in the optical domain and yielded on average 20 points within $\Gamma$. Each single-burst spectrum was averaged 10 times; therefore each $f_{rep}$ step was acquired in 13 s. There was a dead time between two consecutive $f_{rep}$ steps of 4.5 s caused by the need to change the DDS frequency smoothly to avoid unlocking of the comb from the cavity. Thus the entire measurement took 700 s.

The comb $f_{rep}$ and $f_{ceo}$ were continuously monitored during the acquisition with a frequency counter at a rate of 1 Hz and the result is shown in Fig. 8. The 40 $f_{rep}$ steps are clearly visible in panel (a). Because of the comb-cavity locking scheme, $f_{ceo}$ followed a trend opposite to $f_{rep}$ in order to compensate for the $f_{rep}$ increase, as shown in panel (b). The $22^{nd}$ step is enlarged in the insets in both panels. Standard deviations $\sigma$ of $f_{rep}$ and $f_{ceo}$ were equal to 2 mHz and 1.2 kHz, respectively, during each step. This corresponds to a total uncertainty on the comb mode frequency in the optical range equal to 1.7 kHz, mostly caused by the offset frequency uncertainty.

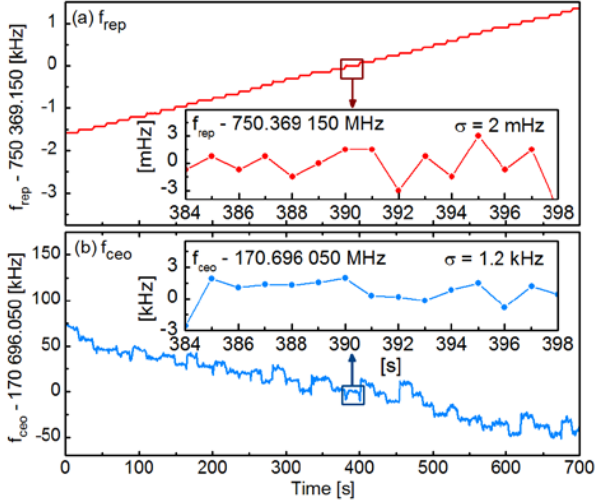

Fig. 8. Experimental $f_{rep}$ (a) and $f_{ceo}$ (b) values measured during the step-by-step acquisition of single-burst interferograms, where the mean value of the 22nd step has been subtracted. The insets show zooms of the 22nd step together with the standard deviation.

The single-burst spectrum measured with the first $f_{rep}$ value was used to reduce the uncertainty $\eta$ on $\lambda_{ref}$ below the value determined by the SNR ratio in the spectrum. All single-burst spectra were then processed using the $f_{shift}^{0}$ shift to correct the FTS scale. When the residual ILS distortion for the absorption lines at the edges of the comb spectral range was found to be larger than the noise on the baseline, the combination of the zero-padding method and the $f_{shift}$ shift was used instead. Each single-burst spectrum was normalized to a background spectrum measured while the cavity was filled with pure $N_2$ at the same pressure. Since the background does not contain any narrow spectral features, a single-burst spectrum taken at the first $f_{rep}$ value (averaged 10 times) was sufficient to fully resolve it. For normalization of spectra taken at other $f_{rep}$ values this background was linearly interpolated at the comb frequencies. The single-burst spectra measured with different $f_{rep}$ values were interleaved and the relative baseline of each step was corrected to compensate for the drift of the laser envelope from step to step. Finally, the baseline of the interleaved spectrum was corrected for the remaining etalons by fitting a 4th order polynomial function and low frequency sine wave functions together with the transmission spectrum.

## 4. Results

### 4.1. Broadband $CO_2$ spectra and sensitivity

Figure 9 shows the spectrum of the $3v_1+v_3$ band of 1000(2) ppm of $CO_2$ diluted in $N_2$ at 26.3(1.3) Torr obtained from interleaved single-burst spectra [panel (b), black curve], compared to a model spectrum (red, inverted) calculated using the cavity-enhanced transmission function [11] based on Voigt absorption profiles with line parameters from the HITRAN database [13], and the experimentally determined cavity finesse [panel (a)]. The nominal resolution of the FTS was 750 MHz while the average FWHM of the absorption lines is around 390 MHz, yet no distortion induced by the ILS is visible. The lack of ringing around the absorption lines is even more evident in the enlargement of the range covering two lines of the $3v_1+v_3$ band, chosen for their high SNR and proximity to one of the PDH locking points, namely the R14e and the R16e lines, plotted in black in panel (c). For comparison, a spectrum that would be obtained with conventional FTIR with the same nominal resolution is plotted in blue (vertically and horizontally offset for clarity). This spectrum was simulated by convolving the measured $CO_2$ spectrum with a sinc function described by Eq. (1) for a nominal resolution of 750 MHz, which causes reduction of the line contrast, broadening of the line width, and ringing on the line wings. None of these effects was present in the experimental spectrum.

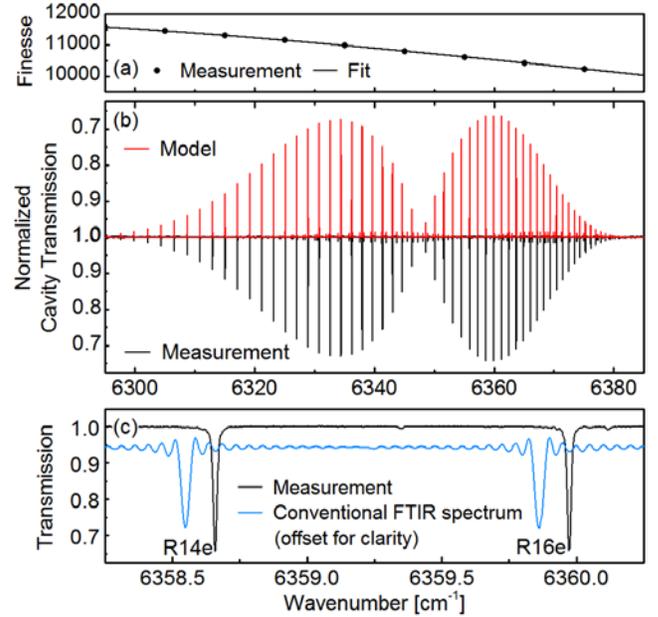

Fig. 9. Cavity-enhanced spectrum of the $3v_1+v_3$ band of $CO_2$ at 26.3 Torr obtained from single-burst interferograms. (a) Cavity finesse measured by cavity ring-down when the cavity was filled with $N_2$ (markers) together with a third order polynomial fit (curve). (b) Normalized cavity transmission spectrum of 1000 ppm of $CO_2$ diluted in $N_2$ at 26.3 Torr obtained from 40 interleaved single-burst spectra (black) plotted together with a simulation based on the cavity enhanced model [11] and the HITRAN parameters [13] (red, inverted). (c) Enlargement of the spectral region containing the R14e and R16e $CO_2$ lines (black) and a simulation of the spectrum that would be obtained with conventional FTIR spectroscopy and the same nominal resolution (750 MHz, blue, horizontally and vertically offset for clarity). The two weak peaks at 6359.34 and 6360.11 $cm^{-1}$ are $CO_2$ lines.

The spectrum spans 90 $cm^{-1}$ between 6295 and 6385 $cm^{-1}$ and contains 144 000 spectral elements spaced by $6.25 \times 10^{-4}$ $cm^{-1}$. The standard deviation of the noise on the baseline near the two lines shown in panel (c) is $1.2 \times 10^{-3}$. Therefore, the noise equivalent absorption (NEA) is equal to $2.2 \times 10^{-9}$ $cm^{-1}$ for an acquisition time of 700 s, and the figure of merit, defined as the NEA multiplied by the square root of the ratio of the total acquisition time and the number of spectral elements, is equal to $1.5 \times 10^{-10}$ $cm^{-1} Hz^{-1/2}$.

### 4.2. Residual ILS distortion

According to Eq. (22), the absence of ILS-induced distortion on the R16e line centered at 6360 $cm^{-1}$ ($n_{abs} = 2.54 \times 10^5$), for which the SNR is 280, implies that the cw reference laser wavelength has been found with an accuracy below $\eta_{lim} = 1.4 \times 10^{-8}$. Figure 10(a)-(b) investigates the ILS distortion of the R16e line when the error $\eta$ is deliberately made larger than this. The undistorted line corresponding to $|\eta| < \eta_{lim}$, same as in Fig. 9(c), is shown by black markers ($\alpha$) in panel (a). The blue and red curves show the same line analyzed using $\lambda'_{ref}$ with $\eta = -5 \times 10^{-8}$ ($\beta$, blue curve) and $\eta = 1 \times 10^{-7}$ ($\gamma$, red curve) with respect to the value used to obtain the ILS-free line, $\lambda_{ref}$ = 632.99115 nm. To quantitatively assess the distortion amplitude, panel (b) shows the differences between curves (β) and (α), and (γ) and (α). As predicted by Eq. (21) and by simulations shown in Fig. 4, the ILS distortion has odd symmetry with respect to the line center and reaches a maximum at a frequency $f_{rep}$ away from the line center. Moreover, the amplitude of these maxima changes sign and increases with η. The dotted horizontal lines in panel (b) indicate the ILS distortion amplitudes calculated using Eq. (21) for curves (β) and (γ), equal to $4.3 \times 10^{-3}$ and $8.6 \times 10^{-3}$, respectively, which agree well with the experimental data in both cases. To assess the center frequency shift caused by η we fit the cavity-enhanced model to the three curves, similarly to what was done to obtain data shown in Fig. 6. The $\Gamma'/f_{rep}$ ratio for this line is 0.52 and we find from the fit that the shift is equal to $-2\eta \times 10^7$ MHz. This implies that the uncertainty of the center frequency for the ILS-free line with $\eta_{lim}$ = $1.4 \times 10^{-8}$ is at most equal to 0.28 MHz.

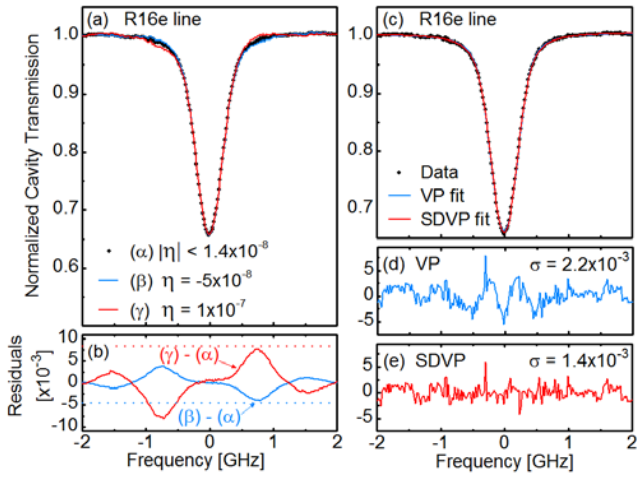

Fig. 10. Spectra of the R16e CO₂ line at 6359.9673 cm⁻¹ at a pressure of 26.3 Torr and temperature of 296 K. (a) Spectra obtained for different values of error on the cw reference laser wavelength: |η| < 1.4×10⁻⁸ [α, black markers, same as in Fig. 9(c)], η = -5×10⁻⁸ (β, red curve), and η = 1×10⁻⁷ (γ, blue curve). (b) Residua between the distorted lines and the undistorted line. The dotted horizontal lines indicate the maximum of the ILS distortion calculated using Eq. (21). (c) Undistorted spectrum [black markers, same as in panel (a)] together with a fit of the Voigt profile (VP, blue curve) and the speed-dependent Voigt profile (SDVP, red curve). Residuals of the fits are shown in (d) for the VP and in (e) for the SDVP.

## 4.3. Speed-dependent effects on Voigt profile

The undistorted R16e CO₂ transition is plotted again with black markers in Fig. 10(c). The nearly overlapping curves show fits to this line based on the cavity-enhanced model [11] with the Voigt profile (VP, blue curve) and with the speed-dependent Voigt profile [14] (SDVP, red curve). The fitting parameters of the VP are the transition frequency, the line intensity, the Lorentzian width of the transition, the comb-cavity offset (induced by the cavity dispersion [11]) and a linear baseline. The Doppler width is fixed to 354 MHz, the pressure shift to - 6.7 MHz (based on the value of the pressure shift by N₂ of -0.0065(7) cm⁻¹/atm from Gamache et al. [15]) and the finesse to 10500. The residual of the VP fit is shown in panel (d). The characteristic (inverted) w-shaped residual induced by the inaccuracy of the VP at this pressure is clearly visible, demonstrating the need to consider more advanced absorption profiles. Panel (e) shows the residual of a fit of the SDVP, where the reduced speed-dependent collisional width was introduced using quadratic approximation given by Priem et al. [16]: $B_w(x) = 1 + a_w(x^2 - 3/2)$, where x is the reduced absorber speed and

$a_w$ is a dimensionless fitting parameter. The small pressure shift was assumed to be independent of the absorber speed. The improvement of the fit within the line core is evident. The standard deviations of the residuals of the VP and SDVP evaluated between -1 and 1 GHz, indicated in their respective panels, show an improvement of the SDVP residual by 36% compared to the VP.

The spectroscopic parameters of the R16e line retrieved from the fits of the VP and the SDVP are summarized in Table 1 and compared to the parameters from the HITRAN database [13] and from Li et al. [17]. The statistic uncertainties of the fitted parameters are indicated in parentheses, while the systematic uncertainties are listed in a separate row. Not listed in the table are the comb-cavity offset equal to 470(95) Hz and 460(70) Hz for the VP and SDVP fits, respectively, as well as the $a_w$ parameter equal to 0.129(7) for the SDVP. The fitted transition frequency agrees well with the HITRAN value. In fact, already at this SNR the precision of our measurement is better than that reported in HITRAN, even taking into account the combined effect of the uncertainty on the pressure shift (0.7 MHz) and that caused by the uncertainty on $\lambda_{ref}$ (0.28 MHz). The retrieved line intensity agrees with the intensity provided in HITRAN within the systematic uncertainty given by the combined effect of the pressure, concentration and cavity finesse uncertainties. The fitted value of the Lorentzian line width of the VP agrees within the systematic uncertainty with the value calculated using the broadening coefficient for N₂ for the same CO₂ line measured by Li et al. ([17], indicated with a star). The systematic uncertainty is given by the precision of the pressure and temperature measurement. The Lorentzian line width found from the SDVP fit is larger, since this parameter does not have to compensate for the speed-dependent line narrowing neglected in the VP.

Table 1. Vacuum transition frequency, line intensity and Lorentzian FWHM, $\Gamma_L$, at 26.3(6) Torr of the R16e CO₂ line retrieved from fits based on the VP and SDVP compared with the parameters from the HITRAN database [13] and from Li et al. [17] at 296(3) K.

| R16e line | Vacuum line position [MHz] | Intensity [cm⁻¹/(mol. cm⁻²)] | $\Gamma_L$ [MHz] |
|---|---|---|---|
| VP (this work) | 190667024.1(5) | 1.650(5)×10⁻²³ | 144(2) |
| SDVP (this work) | 190667024.2(7) | 1.688(5)×10⁻²³ | 163(2) |
| Systematic uncertainty | 0.8 | 0.09 | 7 |
| HITRAN and Li et al. * | 190667023(3) | 1.744(35)×10⁻²³ | 151(5)* |

## 5. Conclusions

Optical frequency comb Fourier transform spectroscopy with sub-nominal resolution allows precise measurement of the power of the individual comb modes interacting with molecules and yields absorption spectra with frequency resolution and precision given by the comb rather than the FTS. It relies on the measurement of single-burst interferograms with nominal resolution precisely matched to the repetition rate of the comb and matching of the FTS sampling points to the comb mode frequencies in post-processing. Any residual mismatch between the FTS and OFC scales induces a characteristic ILS distortion with odd symmetry with respect to the line center and a maximum at $f_{rep}$ away from the line center, with amplitude proportional to the frequency mismatch. This ILS distortion can be reduced below the noise level by iteratively adjusting the cw reference laser wavelength down to the precision $\eta_{lim}$ given by the inverse of the product of the SNR in the spectrum and the absorbed comb mode number, $n_{abs}$.

Importantly, the parameters of absorption lines much narrower than $f_{rep}$ are unaffected by the residual ILS distortion. Thus, to fully benefit from the sub-nominal resolution method, $f_{rep}$ should be chosen larger than the line width of the measured absorption lines, which can be easily achieved in practice using a filter cavity.

The sub-nominal resolution method is similar to two approaches used in dual comb spectroscopy. The first is the 'coherent averaging' technique [18] developed to allow the averaging of successive single-burst interferograms in the time domain to reduce the amount of data. In this method, reproducible sampling of the interferogram is obtained by setting the repetition rates and the difference of the offset frequencies of the interfering combs equal to integer multiples of the difference of the repetition rates, which after FFT yields sampling points exactly at the comb mode frequencies. The second method is the 'super resolution' technique developed initially for the THz range [19] and later used in the near-infrared range [20]. The lack of $f_{ceo}$ in the THz combs removes the need to correct the FTS offset frequency. Moreover, because of the rather low SNR in the spectra measured in both frequency ranges, precise matching of the FTS and OFC scales was not necessary and the effects described in this work were not observed. Our approach allows obtaining ILS-free spectra with a much higher signal-to-noise ratio and can be implemented in all FTS systems, yielding resolutions orders of magnitude better than most accurate FTIR spectrometers [3] using a compact interferometer ($\Delta_{max}$ of tens of cm) and in a much shorter acquisition time (seconds or minutes instead of hours or days). As in conventional FTIR, the temperature and pressure need to be stable during the acquisition process; but contrary to FTIR, vacuum conditions are not required since the absolute frequency scale is set by the light source and not by the spectrometer. Moreover, all uncertainties on the FTS scale can be compensated for in post-processing.

Using the sub-nominal method we measured an ILS-free low-pressure spectrum of the entire $3\nu_1 + \nu_3$ band of $CO_2$ spanning 90 cm$^{-1}$ with optical sampling step of $6.25 \times 10^{-4}$ cm$^{-1}$ in 700 s with a SNR of 280 on the strongest lines. This SNR was sufficient to demonstrate - for the first time with comb-based spectroscopy - that the speed-dependent Voigt profile improves the fit quality for low-pressure absorption lines. This profile is fully compatible with the Hartmann-Tran profile (HTP) [21] that has been recently recommended by the International Union of Pure and Applied Chemistry [22] for use in future spectroscopic databases [22]. Improving the SNR by longer averaging will further improve the precision of the fitted parameters.

The sub-nominal technique enables systematic measurements of entire absorption bands with precision needed for retrieval of line parameters beyond the Voigt profile [22, 23] in acquisition times of the order of minutes, yielding high fidelity spectroscopic data with little influence of systematic errors. The comb-based FTS is thus an ideal tool for high-accuracy broadband measurements of parameters of millions of lines included in HITRAN and other databases, which would take years to measure with other techniques.

**Acknowledgments:** The authors thank Arkadiusz Tkacz for optimizing the FTS acquisition process and the anonymous reviewer for his/her comments that helped us significantly improve the clarity of the manuscript.

**Funding:** This work was supported by the Swedish Research Council (621-2012-3650), the Swedish Foundation for Strategic Research (ICA12-0031), the Knut and Alice Wallenberg Foundation (KAW 2015.0159), and the Polish National Science Centre (DEC-2012/05/D/ST2/01914).